\documentclass[twocolumn,aps,epsf,superscriptaddress]{revtex4-2}

\usepackage{amsmath}
\usepackage{amsfonts}
\usepackage{amssymb}
\usepackage{epsfig}
\usepackage{epsf}
\usepackage{array}
\usepackage{color}
\usepackage{ulem}
\usepackage{units}
\usepackage{hyperref}

\usepackage{setspace}

\renewcommand{\bibsection}{\section*{R\lowercase{eferences}}}

\begin{document}

\title{Berry curvature and field-induced intrinsic anomalous Hall effect in an antiferromagnet FeTe}
\altaffiliation{Copyright  notice: This  manuscript  has  been  authored  by  UT-Battelle, LLC under Contract No. DE-AC05-00OR22725 with the U.S.  Department  of  Energy.   
The  United  States  Government  retains  and  the  publisher,  by  accepting  the  article  for  publication, 
acknowledges  that  the  United  States  Government  retains  a  non-exclusive, paid-up, irrevocable, world-wide license to publish 
or reproduce the published form of this manuscript, or allow others to do so, for United States Government purposes.  
The Department of Energy will provide public access to these results of federally sponsored  research  in  accordance  with  the  DOE  Public  Access  Plan 
(http://energy.gov/downloads/doe-public-access-plan)}. 
%
\author{Satoshi Okamoto}
\altaffiliation{Correspondence and requests for materials should be addressed to okapon@ornl.gov}
\affiliation{Materials Science and Technology Division, Oak Ridge National Laboratory, Oak Ridge, Tennessee 37831, USA}
\author{Adriana Moreo}
\affiliation{Department of Physics and Astronomy, The University of Tennessee, Knoxville, Tennessee 37996, USA}
\affiliation{Materials Science and Technology Division, Oak Ridge National Laboratory, Oak Ridge, Tennessee 37831, USA}
\author{Naoto Nagaosa}
\affiliation{RIKEN Center for Emergent Matter Science (CEMS), Wako, Saitama 351-0198, Japan}
\affiliation{Fundamental Quantum Science Program (FQSP), TRIP Headquarters, RIKEN, Wako, Saitama 351-0198, Japan}
\author{Stuart S. P. Parkin}
\affiliation{Max Planck Institute of Microstructure Physics, Halle(Saale) 06120, Germany}

\begin{abstract}
Berry curvature is ubiquitous in condensed matter physics and materials science. 
Its main consequence is the intrinsic anomalous Hall effect (AHE) in magnetic materials and plays a pivotal role in  
spintronic applications and quantum technologies.  
Here, we present a theoretical study of the intrinsic AHE in tetragonal FeTe, a semimetallic van der Waals antiferromagnet with 
compensated magnetic ordering at low temperatures. 
Using a realistic spin-fermion model, we demonstrate that FeTe exhibits a large Berry-curvature-driven AHE under an applied magnetic field. 
Our calculations reveal that the Hall conductivity of this compound is extremely sensitive to temperature and field strength and even exhibits sign reversal, 
highlighting FeTe as a prototypical platform where magnetism and topology combine to produce robust intrinsic Hall responses. 
This work establishes FeTe as a promising candidate for exploring quantum transport in low-dimensional correlated systems.
We also discuss the implications for recent experimental results of the AHE and ordinary Hall effect reported for FeTe. 
\end{abstract}
\maketitle

\section*{I\lowercase{ntroduction}}

The Hall effect---the generation of a transverse voltage upon applying a current and an out-of-plane magnetic field---is a vital probe of electronic, 
topological and magnetic properties in materials. 
It encompasses two primary contributions: the ordinary Hall effect (OHE), which arises from the Lorentz force on charge carriers, and the anomalous Hall effect (AHE) \cite{Nagaosa2010}, 
a more complex phenomenon originating in magnetic systems and typically related to the material's net magnetization. 
Its origin is split into extrinsic effects arising from scattering mechanisms induced by magnetization and intrinsic effects rooted in the Berry curvature of the electronic band structure. 
Studies of the AHE have proven instrumental in exploring topological and quantum phenomena in a wide range of materials \cite{Fang2003,Chen2014,Kang2019,Zhou2022}.

Antiferromagnetic (AFM) materials, with their compensated spin sublattices, are potential candidate materials for future ultrafast, low-power spin-based logic devices. 
They pose challenges, however, for the study of the AHE which requires time reversal symmetry breaking \cite{Smejkal2022,Shindou2001,Balz2018,Peng2020,Liu2025}. 
Recent studies have demonstrated an intrinsic AHE in geometrically frustrated noncollinear antiferromagnets \cite{Nakatsuji2015,Nayak2015} and, 
more recently, in collinear antiferromagnets \cite{Zhu2025}. 
Antiferromagnetic van der Waals (vdW) materials have shown great potential for applications due to their rich topological and quantum properties \cite{Deng2020,Chen2024}. 
Furthermore, they make readily accessible layers with thicknesses at the atomic limit \cite{Bong2017,Huang2017,Deng2018}. Few-layer insulating Cr{\it X}$_3$ ({\it X} = I, Cl) \cite{Huang2017,McGuire2017a}, 
MnBi$_2$Te$_4$ \cite{Li2019,Wu2019}, and {\it Y}PS$_3$ ({\it Y}=Mn, Fe, Ni) \cite{Joy1992,Lee2016}, as well as the 
conducting FeTe \cite{Kang2020}, CrTe$_3$ \cite{McGuire2017b}, and GdTe$_3$ \cite{Lei2020}, have been identified with collinear AFM orders. 
The collinear order exhibits an ideal compensation of opposite spin lattices with, therefore, negligible net magnetization. 
This not only heightens the advantage for devices of minimal stray magnetic fields, but also additionally suppresses any extrinsic mechanisms of AHE that 
requires a net spin polarization for asymmetric scattering of the conduction electrons. 
For example, when one considers the spin cluster scattering in the presence of scalar spin chirality, skew scattering occurs in the absence of any magnetization~\cite{Ishizuka2018}.
As the origin of large AHE in most AFMs, an intrinsic mechanism exists when a specific symmetry is broken, and, therefore, 
is the most likely mechanism to generate an AHE for the collinear AFM cases \cite{Zhou2022,Shao2020,Li2023}. 

Tetragonal FeTe maintains frustration induced by direct and indirect Fe-Fe hopping \cite{Yin2011}, 
allowing the possibility of observing an AHE in collinear vdW AFMs. 
Meanwhile, FeTe is a semimetal, which indicates that a large AHE could emerge from the Berry curvature around nodal lines or points \cite{Manna2018}. 
In this work, we develop such a theoretical framework based on a realistic spin-fermion model. 
The model couples itinerant Fe $d$-electrons with localized magnetic moments, with its parameters derived from density functional theory (DFT) calculations and Wannier tight-binding constructions. 
This approach allows us to capture the evolution of electronic states across the AFM transition and to incorporate the effects of spin-orbit coupling that generates Berry curvature. 

\begin{figure*}
\begin{center}
\includegraphics[width=1.7\columnwidth, clip]{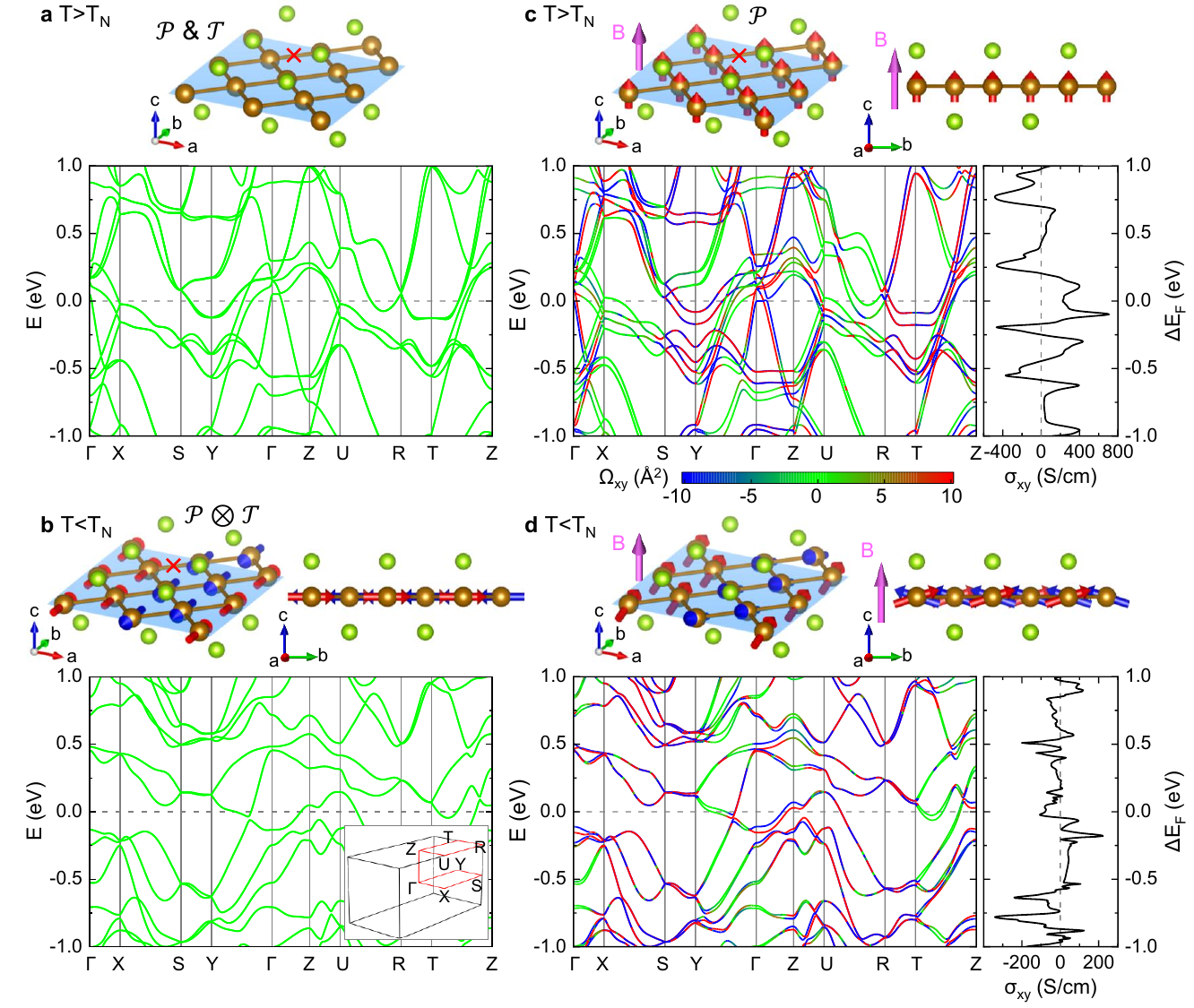}
\caption{Interconnection between magnetism, band structure, and anomalous Hall conductivity. 
{\bf a}. Above $T_{\rm N}$ without a magnetic field. 
Because of the spatial inversion ($\cal P$) symmetry and the time-reversal ($\cal T$) symmetry, all bands are twofold degenerate, i.e., Kramer's doublet. 
{\bf b}. Below $T_{\rm N}$ without a magnetic field. 
While the bicollinear AFM ordering breaks $\cal T$ symmetry, the combined ${\cal P} \otimes {\cal T}$ symmetry remains, protecting the twofold degeneracy. 
Accordingly, there is no Berry curvature.  
{\bf c}. Above $T_{\rm N}$ with a magnetic field. 
$\cal T$ symmetry is broken by an external magnetic field, resulting in nonzero Berry curvature and anomalous Hall conductivity $\sigma_{xy}$. 
{\bf d}. Below $T_{\rm N}$ with a magnetic field. 
${\cal P} \otimes {\cal T}$ symmetry is broken by an external magnetic field, resulting in nonzero Berry curvature and anomalous Hall conductivity $\sigma_{xy}$. 
The band structure, Berry curvature, and $\sigma_{xy}$ are computed using the spin-fermion model at $T=1.2\,T_{\rm N}$ ($0.4 \,T_{\rm N}$) for {\bf a} and {\bf c} 
({\bf b}. and {\bf d}). 
A magnetic field $B=10$~Tesla is applied along the $z$ direction for {\bf c} and {\bf d}. 
In all band plots, the reduced Brillouin zone is used, as shown in the inset of {\bf b}. 
Inversion centers are indicated by red crosses in the structure models. 
Nominal Fermi level $E_{\rm F}$ associated with 6 electrons per Fe site is set to zero. 
$\sigma_{xy}$ is plotted as a function of the Femi level $\Delta E_{\rm F}$ shifted from the nominal value.
Structure models are produced using the VESTA software \cite{VESTA}. 
}
\label{fig:MagBerryAHC}
\end{center}
\end{figure*}

The main purpose of this study is therefore to provide a microscopic and quantitative understanding of the intrinsic AHE in a collinear AFM material. 
By computing the anomalous Hall conductivity from the Berry curvature distribution in the Brillouin zone,
we demonstrate that the large AHE in FeTe is intrinsically driven and directly linked to the evolution of the bicollinear AFM order under finite magnetic fields. 
In particular, our calculations reveal a temperature- and doping-dependent crossover in the sign of the Hall conductivity and 
highlight the critical role played by band topology in governing the Hall response of FeTe, 
{
similar to well-known  SrRuO$_3$ \cite{Fang2003}}. 
This establishes FeTe as a prototypical vdW AFM platform where magnetism and topology cooperate to yield strong intrinsic Hall responses. 
The framework presented here may also be extended to other correlated vdW magnets, offering a general route for exploring Berry-curvature-driven transport phenomena 
in low-dimensional systems.

\section*{R\lowercase{esults}}
\label{sec:results}

Here we describe our theoretical investigation on the Berry curvature contribution to the anomalous Hall conductivity (AHC) in FeTe. 
Our analysis is based on a realistic spin-fermion model, 
which consists of itinerant electrons exchange coupled with localized spins \cite{Liang2012}. 
This approach allows one to investigate the electronic property at finite temperature in the presence of an applied magnetic field. 
The Hamiltonian is given by $H_{\rm sf}=H_d + H_S + H_{d-S}$. 
Here, $H_d$ and $H_S$ describe itinerant Fe $d$ electrons and localized Fe moments, respectively, and $H_{d-S}$ describes the exchange coupling between 
itinerant electrons and localized moments. 
$H_d$ is derived from DFT calculations using the experimental structure at 80~K reported in Ref.~\cite{Li2009}
and Wannier tight-binding constructions \cite{Pizzi2020}. 
For their explicit forms and detailed derivations, see Methods and the Supplementary Information. 

{
We apply a static mean-field approximation to $H_S$ to obtain a magnetic state at a given temperature and an applied magnetic field.
The band structure of Fe $d$ electrons is then determined from $H_d + H_{d-S}$.
Here, the Fermi level $E_{\rm F}$ is fixed via 
$N_d= \frac{1}{N} \sum_{\alpha, {\bf k}}f\left(\varepsilon_{\alpha \mathbf{k}} - E_{\rm F}\right)$, with
$N_d=24$ the total number of Fe $d$ electrons in a magnetic unit cell, corresponding to 6 electrons per Fe, 
and $N$ the total number of magnetic unit cells. 
$\varepsilon_{\alpha \mathbf{k}}$ is the energy eigenvalue of band $\alpha$ at momentum $\bf k$, 
and $f\left(\varepsilon\right) = \frac{1}{\{ \exp (\varepsilon/k_{\rm B}T)+1 \}}$, i.e., the Fermi-Dirac distribution function. 
Then, the Berry curvature and AHC are computed at a given temperature and an applied magnetic field. 
Thus,} the AHC is influenced by temperature via two sources, one is the magnetic order parameter, which modifies the band structure, and 
another is the Fermi distribution function. 
For details, see the Methods section.

\begin{figure*}
\begin{center}
\includegraphics[width=1.7\columnwidth, clip]{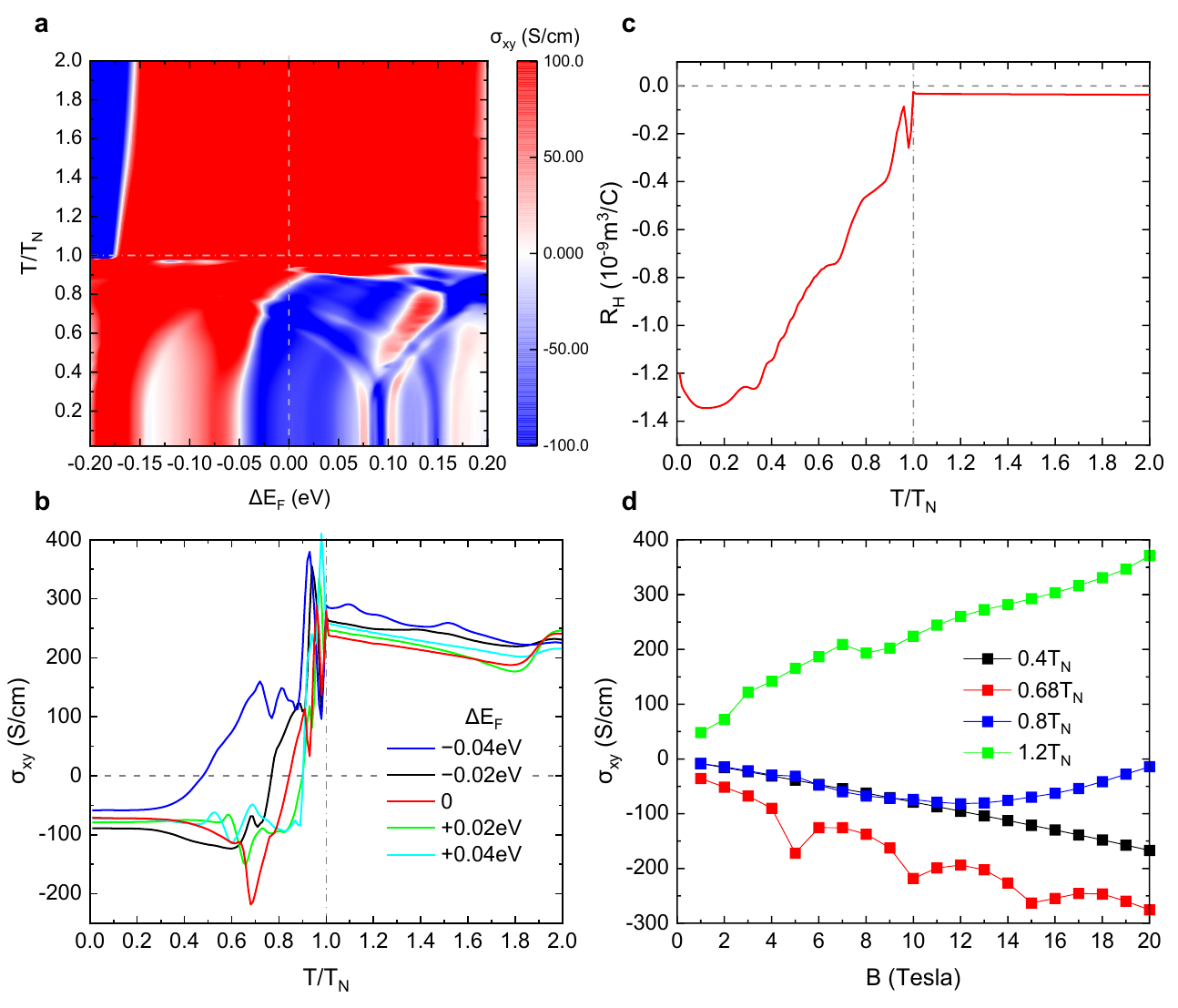}
\caption{Temperature and field evolution of the electronic structure of FeTe. 
{\bf a}. Color map of AHC $\sigma_{xy}$ at $B=10$~Tesla plotted against temperature and the Fermi level $\Delta E_{\rm F}$ shifted 
from the nominal value, corresponding to 6 electrons per Fe site. 
{\bf b}. Temperature dependent anomalous Hall conductivity AHC $\sigma_{xy}$ at selected Fermi energy shifts $\Delta E_{\rm F}$. 
{\bf c}. Temperature dependent ordinary Hall coefficient (OHC) $R_{\rm H}$ computed at the nominal Fermi level, i.e., 
$\Delta E_{\rm F}=0$. 
{\bf d}. Magnetic-field dependent AHC $\sigma_{xy}$ at selected temperatures. 
$\sigma_{xy}$ at $B=10$~Tesla has the negative maximum at $T=0.68 \,T_{\rm N}$ with the nominal Fermi level. 
}
\label{fig:AHC_OHC}
\end{center}
\end{figure*}

\begin{figure*}
\begin{center}
\includegraphics[width=1.8\columnwidth, clip]{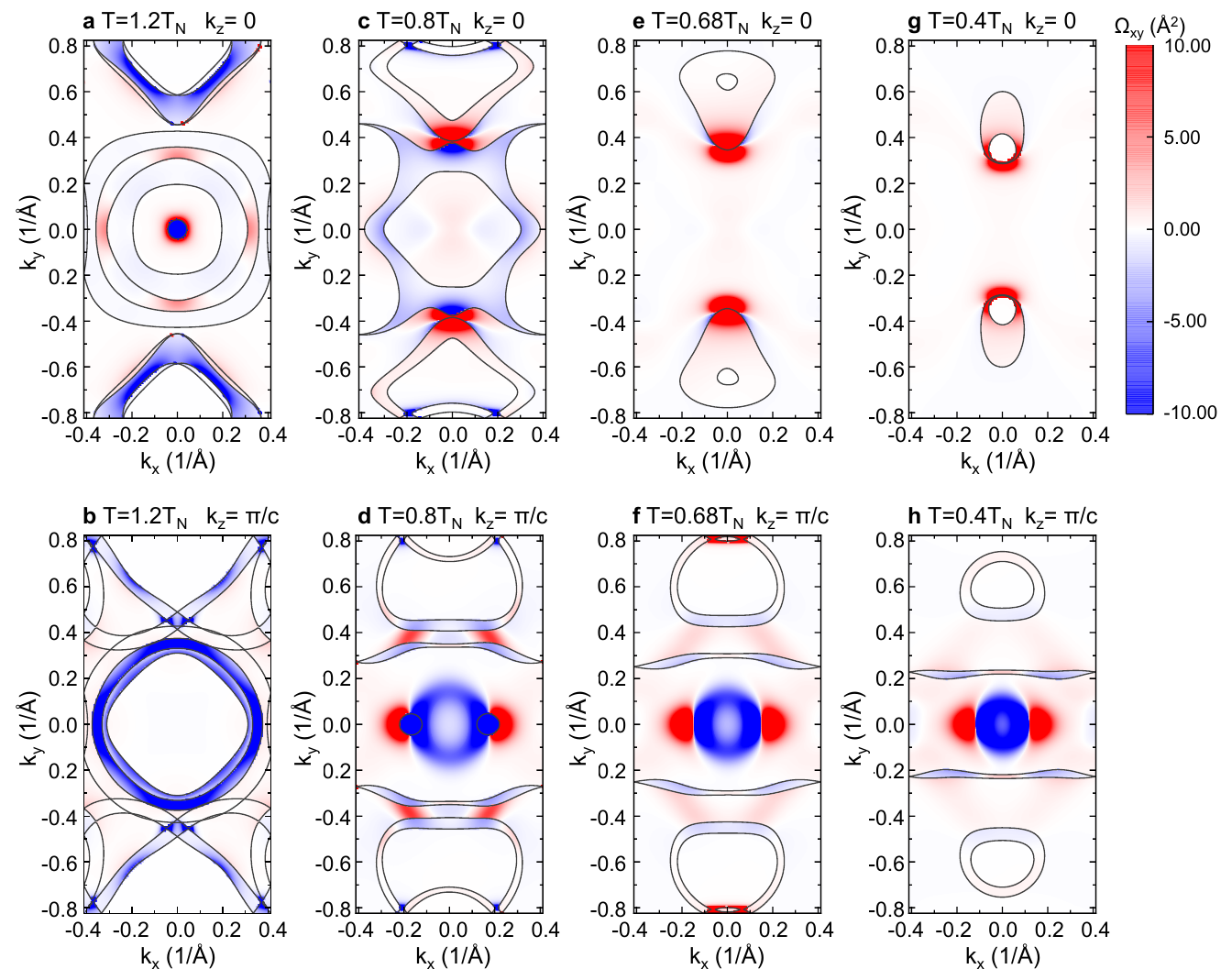}
\caption{Fermi surfaces and Berry curvature ($\Omega_{xy}$) integrated over the occupied states and Fermi surfaces under a finite magnetic field $B = 10$~T. 
{\bf a} and {\bf b}. $T = 1.2\, T_{\mathrm N}$ (above $T_{\mathrm N}$). 
{\bf c} and {\bf d}.  $T = 0.8 \,T_{\mathrm N}$, {\bf e} and {\bf f}. $T = 0.68 \, T_{\mathrm N}$, and 
{\bf g} and {\bf h}. $T = 0.4\, T_{\mathrm N}$(below $T_{\mathrm N}$). 
Panels {\bf a}, {\bf c}, {\bf e}, and {\bf g} show the momentum cuts at $k_z = 0$, 
and panels {\bf b}, {\bf d}, {\bf f}, and {\bf g} show the momentum cut at $k_z = \pi/c$. 
The color bar indicates the magnitude of $\Omega_{xy}$ in the units of {\AA}$^2$. 
$c=6.2517$~{\AA} is the $c$-axis lattice constant \cite{Li2009}.}
\label{fig:FS_B10}
\end{center}
\end{figure*}

\begin{figure*}
\begin{center}
\includegraphics[width=1.7\columnwidth, clip]{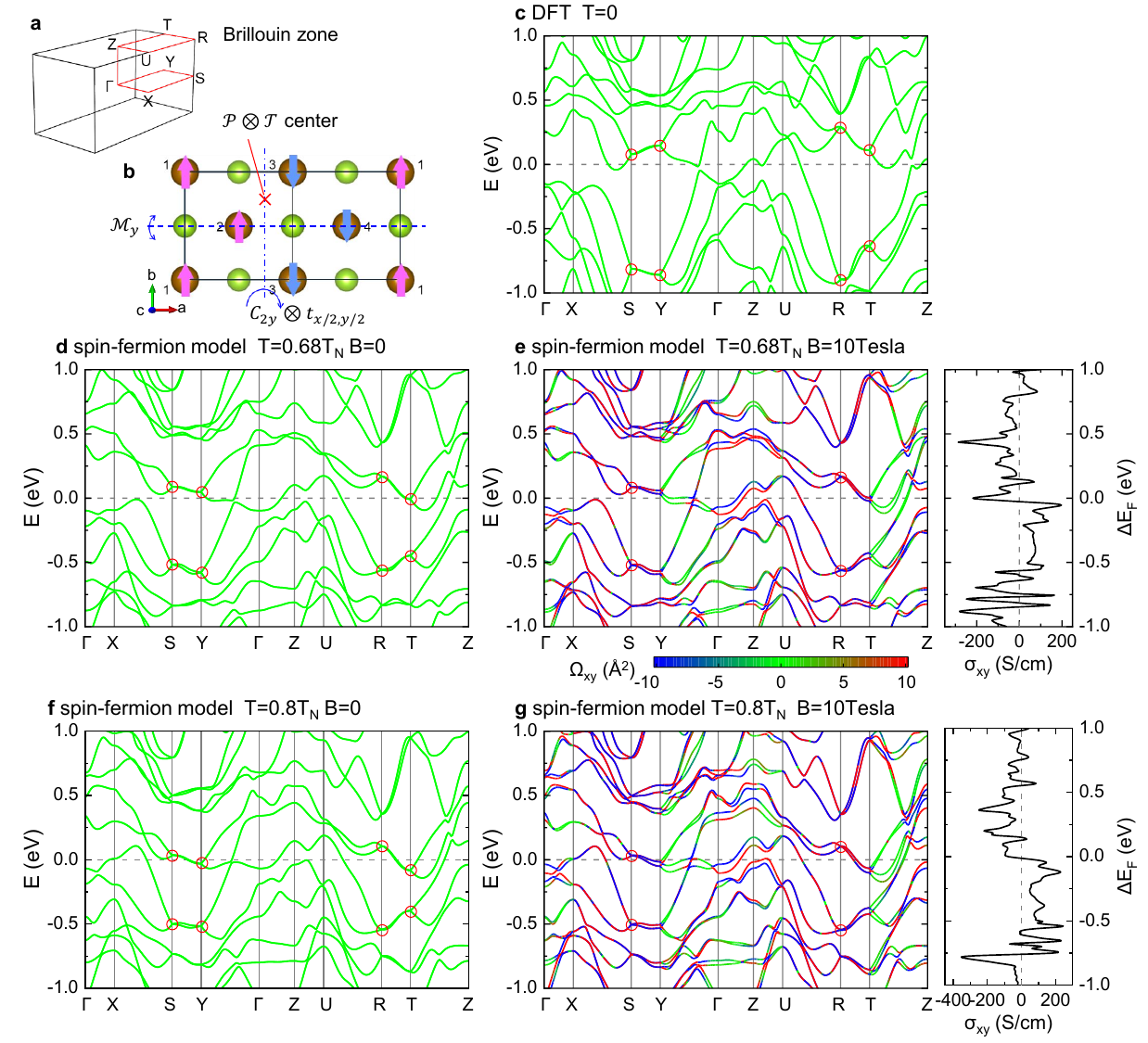}
\caption{Symmetry and band structure of FeTe below N\'eel temperature. 
{\bf a}. Magnetic Brillouin zone. {\bf b}. 
Three symmetries under which bicollinear AFM ordering is invariant: 
${\cal P} \otimes {\cal T}$, mirror reflection with respect to the $y$ plane ${\cal M}_y$, 
and twofold rotation with respect to the $y$ axis followed by half lattice translation along $x$ and $y$ directions ${\cal C} \otimes t_{x/2,y/2}$. 
{\bf c}. DFT band structure with bicollinear AFM ordering at $T=0$. 
{\bf d}, {\bf e}. Band structures deduced from the spin-fermion model at $T=0.68 \,T_{\rm N}$ with $B=0$ and $B=10$~Tesla, respectively. 
{\bf f}, {\bf g}. Band structures deduced from the spin-fermion model at $T=0.8 \,T_{\rm N}$ with $B=0$ and $B=10$~Tesla, respectively. 
With $B=0$, the three symmetries shown in {\bf b} protect band crossings at the $\rm S$, $\rm Y$, $\rm R$, and $\rm T$ points, 
as indicated by the red circles in {\bf d} and {\bf f}. 
With nonzero $B$, these symmetries are broken. 
However, there remain three other symmetries, $\cal P$ followed by half lattice translation along the $t_{x/2}$ axis, 
${\cal C}_{2y}$ followed by half lattice translation along the $y$ direction $t_{y/2}$ and $\cal T$, 
and ${\cal M}_y$ followed by $t_{x/2}$ and $\cal T$, 
which protect band crossings at the $\rm S$ and $\rm R$ points, as indicated by the red circles in {\bf e} and {\bf g}. 
In {\bf c}--{\bf g}, energy is measured from the nominal Fermi level $E_{\rm F}$, corresponding to 6 electrons per Fe site. 
In {\bf e} and {\bf g}, band-dependent Berry curvature is indicated by the color code, with 
the AHC $\sigma_{xy}$ plotted against the Fermi level shift from the nominal value. }
\label{fig:band_details}
\end{center}
\end{figure*}

\begin{figure*}
\begin{center}
\includegraphics[width=1.6\columnwidth, clip]{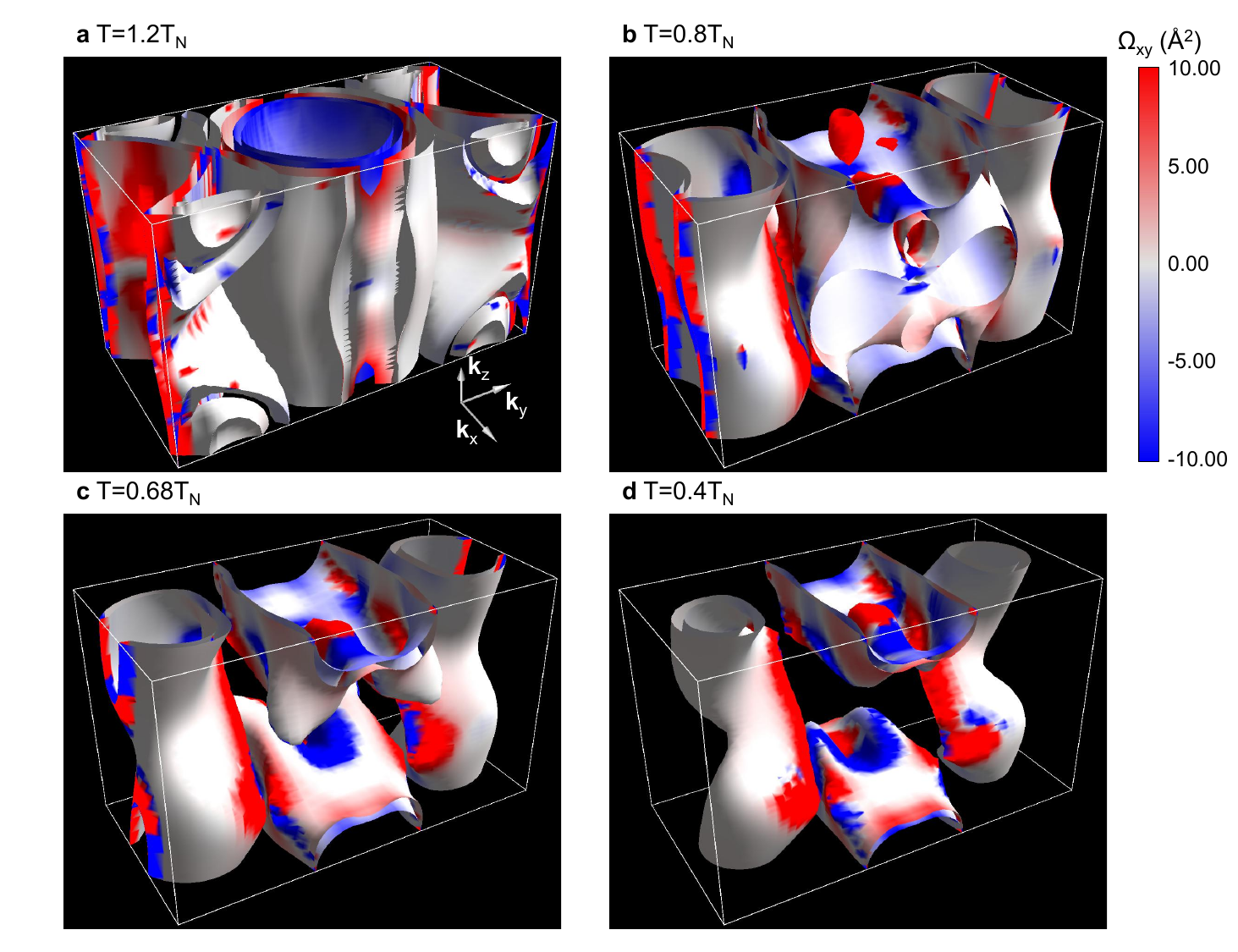}
\caption{Three-dimensional plots of Fermi surfaces with $B=10$~Tesla. 
{\bf a}. $T = 1.2\, T_{\mathrm N}$ (above $T_{\mathrm N}$) , 
{\bf b}. $T = 0.8\, T_{\mathrm N}$, 
{\bf c}. $T = 0.68 \,T_{\mathrm N}$, 
and {\bf d}. $T = 0.4\, T_{\mathrm N}$(below $T_{\mathrm N}$). 
Color map shows the amplitude of Berry curvature on the Fermi surfaces. 
These plots are created using FermiSurfer \cite{Kawamura2019}.}
\label{fig:FS_3D}
\end{center}
\end{figure*}

\begin{figure*}
\begin{center}
\includegraphics[width=1.8\columnwidth, clip]{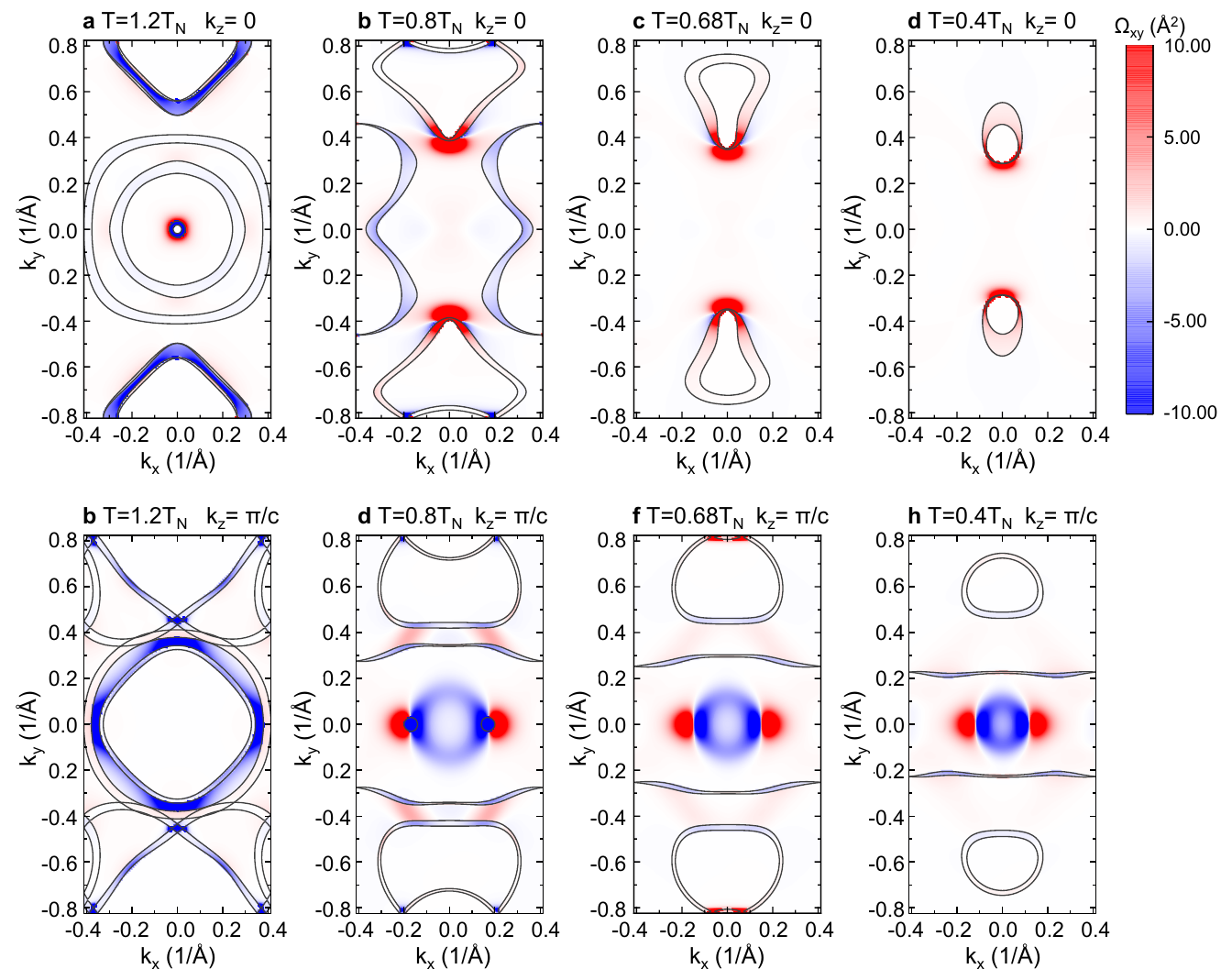}
\caption{{
Same as in Figure~\ref{fig:FS_B10}, but with a different magnetic field $B = 5$~Tesla.} 
}
\label{fig:FS_B5}
\end{center}
\end{figure*}

First, we summarize the symmetry and resulting anomalous Hall responses of FeTe in two temperature regimes, 
above and below the N\'eel temperature, {
$T_{\rm N}=60$~K}, 
in Figure~\ref{fig:MagBerryAHC}. 
Above $T_{\rm N}$ without an external magnetic field, the system has both spatial-inversion ($\cal P$) and time-reversal ($\cal T$) symmetries. 
This enforces the twofold Kramer's degeneracy across all bands (see Fig.~\ref{fig:MagBerryAHC}~{\bf a}) and thus gives rise to zero AHC. 
Similarly, below $T_{\rm N}$ without an external magnetic field, the system has the ${\cal P} \otimes {\cal T}$ combined symmetry, 
enforcing the twofold band degeneracy (see Fig.~\ref{fig:MagBerryAHC}~{\bf b}) and again zero AHC. 
Applying a magnetic field induces a net magnetization (see Fig.~\ref{fig:MagBerryAHC}~{\bf c} and {\bf d}) to break such symmetries, lifting band degeneracies and opening gaps at certain symmetry-protected band crossings, 
depending on the details of the band structure and symmetry protection (Extended Data Fig.~\ref{fig:band_details}). 
This induces Berry curvature in momentum space (Extended Data Fig.~\ref{fig:FS_3D}) and generates a nonzero Berry curvature integral below the Fermi surface, i.e., the AHC. 
It is found that this behavior is more pronounced with increasing magnetic field, 
providing a plausible intrinsic origin for a significant AHC 
{
(compare the Berry curvature plots at different magnetic fields Fig.~\ref{fig:FS_B10} with $B=10$~Tesla
and Extended Data Fig.~\ref{fig:FS_B5} with $B=5$~Tesla)}. 
{
It should be noted that the Berry curvature and AHC are antisymmetric with respect to the B field direction.}

Figure~\ref{fig:AHC_OHC}~{\bf a} shows the color map of calculated AHC $\sigma_{xy}$ as a function of temperature $T$ and 
the shift in the Fermi level $\Delta E_{\rm F}$ under an out-of-plane magnetic field of $B = 10$~Tesla. 
See Methods for the details of the calculations. 
Here, $\Delta E_{\rm F} = 0$ corresponds to the nominal Fermi energy associated with 6 electrons per Fe site,  
thus $\Delta E_{\rm F} > 0$ and $\Delta E_{\rm F} < 0$ represent electron and hole doping, respectively. 
{
In addition to chemical doping, Fermi level could be tuned by using the field effect.}
Across a broad range of $\Delta E_{\rm F}$ values, the AHC exhibits a robust temperature dependence: 
above $T_{\rm N}$, the AHC remains large and positive (in red color), 
while just below $T_{\rm N}$, 
it rapidly switches to a negative value (in blue color) within a narrow temperature range due to the rapid change of band structure involving the emergence of AFM order. 
The AHC crossover temperature, at which AHC equals zero (in white color), is highly sensitive to $\Delta E_{\rm F}$, particularly under hole doping, i.e. $\Delta E_{\rm F}<0$. 
As illustrated in the line profiles along different $\Delta E_{\rm F}$ values (Fig. \ref{fig:AHC_OHC}~{\bf b}), 
even a slight shift of Fermi level toward the hole side significantly reduces the crossover temperature, 
while other characteristics remain largely unchanged. 
Such a sensitivity strongly suggest that the temperature dependence of the AHC originates from the change in the Fermi surface topology. 
This is partly supported by the plots of $\sigma_{xy}$ as a function of applied magnetic field in {
Fig.~\ref{fig:AHC_OHC}~{\bf d}}. 
Here, $\sigma_{xy}$ shows a nearly linear dependence on $B$ at $T=0.4 \,T_{\rm N}$, $0.8 \,T_{\rm N}$, and $1.2 \,T_{\rm N}$, 
while $\sigma_{xy}$ shows a nonlinear behavior at $0.68 \,T_{\rm N}$. 
As shown in Fig.~\ref{fig:AHC_OHC}~{\bf b}, $\sigma_{xy}$ at $B=10$~Tesla shows a pronounced negative peak at $T=0.68 \,T_{\rm N}$. 
Thus, an applied $B$ field controls band dispersions, such as opening a gap at certain band crossings located near $E=0$, 
and thereby induces drastic changes in Berry curvature in this temperature range. 
The AHC is expected to decrease in magnitude with further lowering of the temperature because the Fermi volume decreases as the bicollinear AFM ordering develops. 
{
It is observed that $\sigma_{xy}$ reverses its sign below $T_{\rm N}$, rather than exactly at $T_{\rm N}$. 
As will be detailed later, this behavior is linked to the dominance of positive or negative Berry curvature at different momenta and temperatures.}

{
We also examine the ordinary Hall coefficient (OHC) following Ref.~\cite{Yates2007}. }
Since FeTe with its bicolinear AFM ordering is a semimetal with partial gap opening (see DFT band structure in the Supplementary Information), 
OHC $R_{\rm H}=\rho_{yx}/B$ shows a rather monotonic $T$ dependence, as shown in Fig.~\ref{fig:AHC_OHC}~{\bf c}, 
except for small but sharp kinks near $T_{\rm N}$ induced by the bicollinear AFM ordering.  

Based on the evolution of the Berry curvature at various $B$ values, the field dependence of the AHC is calculated, as shown in 
Fig.~\ref{fig:AHC_OHC}~{\bf d} (see also Extended data Fig.~\ref{fig:AHC_vs_B} for the $\Delta E_{\rm F}$ dependence). 
The magnetic field modulates the band structure by influencing the magnetic ordering through the exchange coupling term $H_{d-S}$, 
which adjusts the interaction between localized spins and itinerant electrons, thereby altering the electron dispersion relation and the AHC. 
The AHC increases approximately linearly with the field above $T_{\rm N}$, while it negatively increases with the field below $T_{\rm N}$. 
Notably, across most of the temperature range below $T_{\rm N}$, the AHC maintains a predominantly linear relationship with the field. 
However, near $T = 0.68 \,T_{\rm N}$, the AHC exhibits a pronounced nonlinear behavior in $B$. 
In other words, the spin canting leads to a non-linear evolution of the Berry curvature near $E_{\rm F}$ and AHC. 

To gain further insight into the relationship between the AHC crossover behavior and the AFM order, 
we show the Berry curvature integral ($\Omega_{xy}$) under $B = 10$~T at different temperatures and momentum planes 
(Fig.~\ref{fig:FS_B10} {\bf a} and {\bf b} for $T = 1.2\, T_{\rm N}$, {\bf c} and {\bf d} for $T = 0.8 \,T_{\rm N}$, 
{\bf e} and {\bf f} for $T=0.68\, T_{\rm N}$, and {\bf g} and {\bf h} for $T=0.4 \,T_{\rm N}$).
Compared to the zero-field results (Extended Data Fig.~\ref{fig:FS_B0}), the Fermi surface topology exhibits a pronounced temperature dependence, 
reflecting the evolution of spin ordering across the AFM transition. 
Above $T_{\rm N}$, negative $\Omega_{xy}$ peaks emerge at the hole pockets near the Brillouin zone center (Fig.~\ref{fig:FS_B10} {\bf a}, {\bf b}), contributing to positive AHC. 
By contrast, below $T_{\rm N}$, positive $\Omega_{xy}$ peaks appear at electron pockets away from the zone center on the $k_z = 0$ plane (Fig.~\ref{fig:FS_B10} {\bf c}), 
leading to a negative AHC. 
Meanwhile, the contribution from the $k_z = \pi/c$ plane is negligibly small, as the positive and negative intensities of $\Omega_{xy}$ cancel out (Fig.~\ref{fig:FS_B10} {\bf d}). 
Here, $c=6.2517$~{\AA} is the $c$-axis lattice constant \cite{Li2009}.
{
As  mentioned previously, 
the sign change of the anomalous Hall effect occurs below $T_{\rm N}$ rather than precisely at $T_{\rm N}$. 
This is due to a gradual shift in the dominant contributions from hole pockets above $T_{\rm N}$ to electron pockets below $T_{\rm N}$. 
Because the magnetic phase transition at $T_{\rm N}$ is second order, the competition between these contributions leads to 
a smooth crossover, rather than an abrupt switch at $T_{\rm N}$.
}
%
{
Furthermore, the negative peak in AHC at $T=0.68\, T_{\rm N}$ is understood as arising from a Lifshitz transition. 
As shown in Extended data Fig.~\ref{fig:TN07}, at $T=0.7\,T_{\rm N}$, there remain tiny hole pockets at $k_x \sim 0.4$~\AA$^{-1}$ and 
$k_y \sim \pm 0.3$~\AA$^{-1}$  in the $k_z=0$ plane with negative Berry curvature, 
while such hole pockets do not exist at $T=0.68\, T_{\rm N}$. 
Thus, the disappearance of hole pockets, which competes with the electron pockets that exhibit a strongly positive Berry curvature
(see Fig.~\ref{fig:FS_B10}~{\bf e} and Extended data Fig.~\ref{fig:TN07}~{\bf e}),
leads to the negative peak in AHC.  
}

\section*{D\lowercase{iscussion}}

Our theoretical investigation based on the spin-fermion model revealed that the band dispersion and Berry curvature in FeTe are highly sensitive 
to both temperature and applied magnetic field, particularly across the bicollinear AFM transition. 
This sensitivity leads to a pronounced temperature dependence of the AHC.  
The OHC also exhibits a characteristic temperature dependence, reflecting changes in the electronic structure induced by the bicollinear AFM transition. 

These theoretical predictions can be compared with recently reported experimental results. 
In Ref.~\cite{Tsukada2011}, Tsukada et al. reported detailed measurements of OHC across a wide temperature range, 
showing positive OHC at $T>T_{\rm N}$ {
($R_{\rm H} \sim +2 \times 10^{-9}$~m$^3$/C at $T\sim 200$~K)} 
and negative OHC at $T<T_{\rm N}$ {
($R_{\rm H} \sim -2.5 \times 10^{-9}$~m$^3$/C at $T\sim 50$~K)}. 
Based on the theoretical prediction of carrier type, equal contributions of holes and electrons in a paramagnetic state 
and electron contributions in a collinear AFM state \cite{Ma2009}, 
the temperature-dependent OHC is attributed to a temperature-dependent carrier mobilities, 
with $\mu_{\rm h} > \mu_{\rm e}$ at $T>T_{\rm N}$ and
$\mu_{\rm h} < \mu_{\rm e}$ at $T<T_{\rm N}$, 
where $\mu_{\rm h}$ and $\mu_{\rm e}$ are hole and electron mobilities, respectively.    

\begin{figure*}
\begin{center}
\includegraphics[width=2\columnwidth, clip]{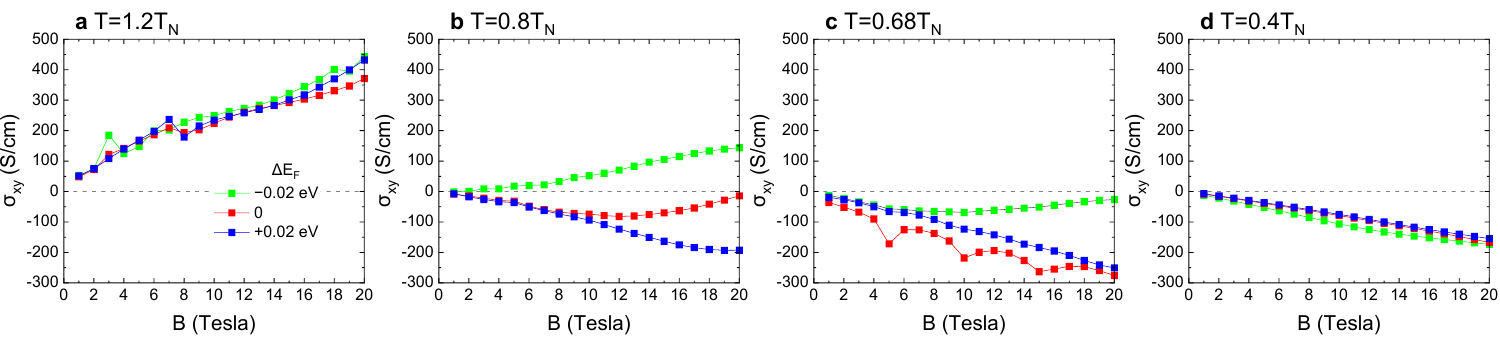}
\caption{Magnetic field-dependent AHC at various temperatures ($T = 1.2\, T_{\mathrm N}$, $0.8 \, T_{\mathrm N}$, $0.68 \, T_{\mathrm N}$ and $0.4\, T_{\mathrm N}$). 
Nonzero carrier doping is also considered by shifting the Fermi level from the nominal value as $\Delta E_{\rm F}=-0.02$~eV, corresponding to hole doping, 
and $\Delta E_{\rm F} = +0.02$~eV, corresponding to electron doping.}
\label{fig:AHC_vs_B}
\end{center}
\end{figure*}

\begin{figure*}
\begin{center}
\includegraphics[width=1.7\columnwidth, clip]{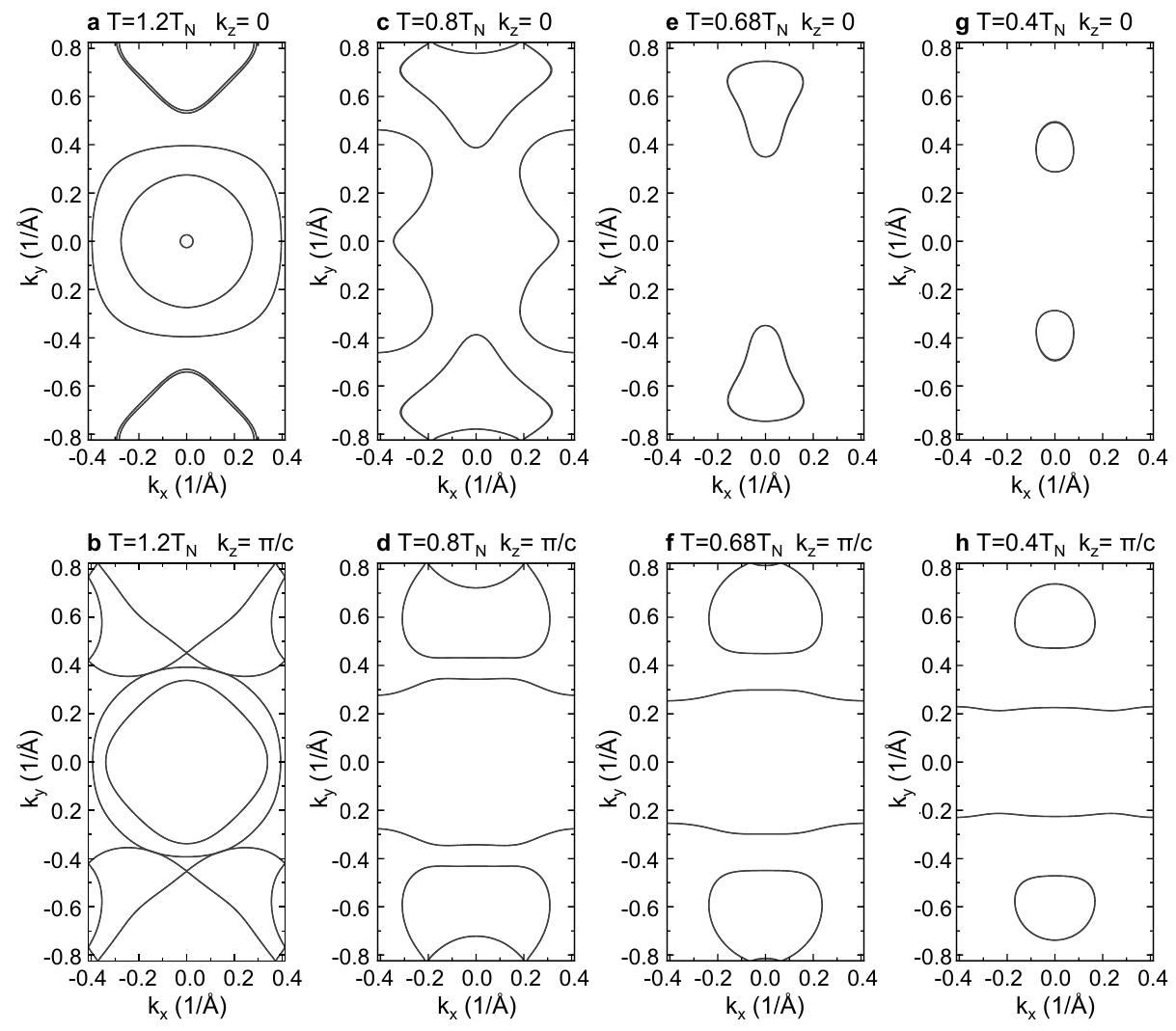}
\caption{Fermi surfaces in zero external field. 
{\bf a} and {\bf b} $T = 1.2 \,T_{\mathrm N}$ (above $T_{\mathrm N}$) , 
{\bf c} and {\bf d}. $T = 0.8 \,T_{\mathrm N}$, 
{\bf e} and {\bf f}. $T = 0.68\, T_{\mathrm N}$, 
and {\bf g} and {\bf h}. $T = 0.4\, T_{\mathrm N}$(below $T_{\mathrm N}$). 
Panels ({\bf a}, {\bf c}, {\bf e}, {\bf g}) show the momentum cut at $k_z = 0$, 
and panels ({\bf b}, {\bf d}, {\bf f}, {\bf g}) show the momentum cut at $k_z = \pi/c$. }
\label{fig:FS_B0}
\end{center}
\end{figure*}

\begin{figure*}
\begin{center}
\includegraphics[width=1.7\columnwidth, clip]{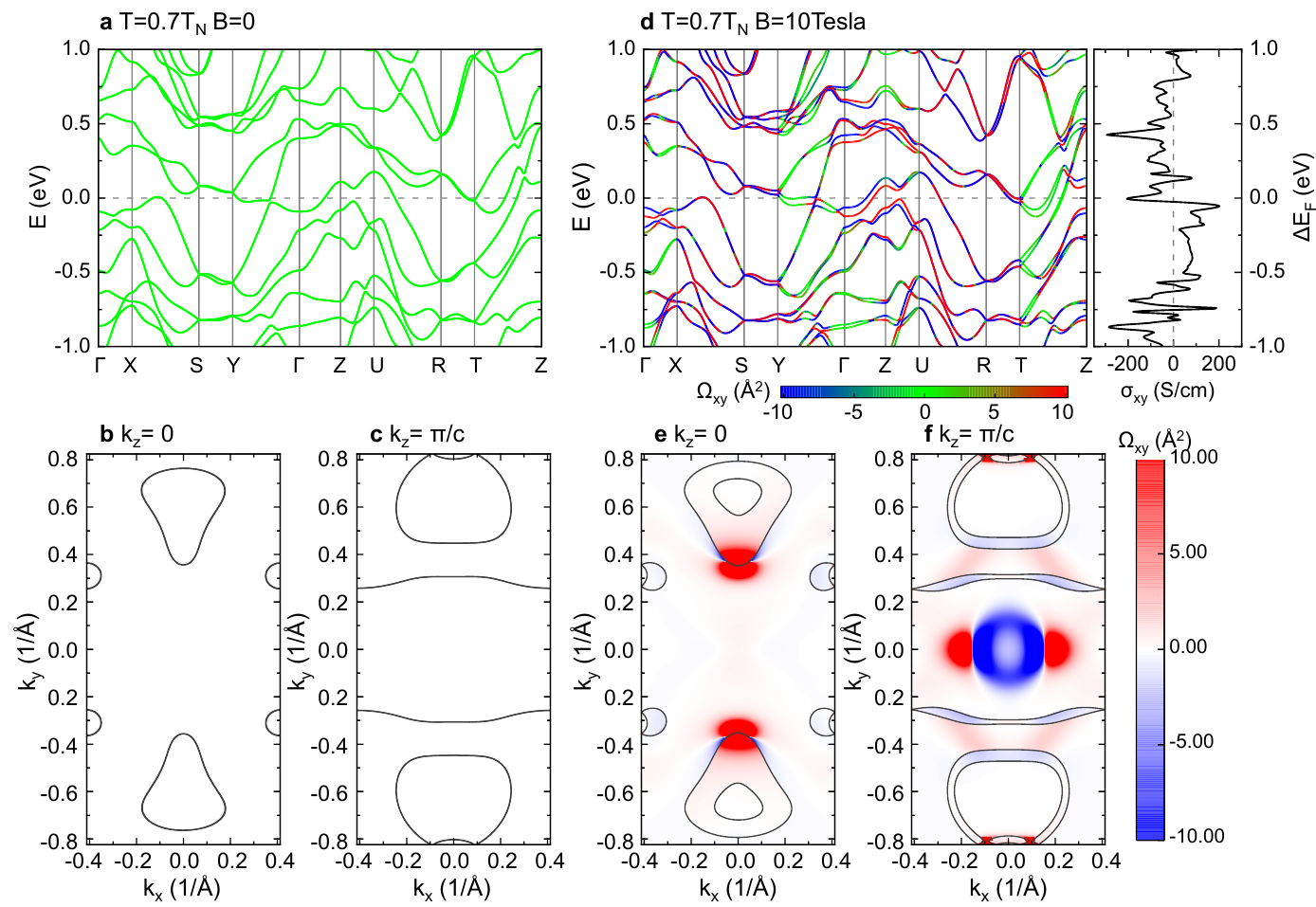}
\caption{Results at $T=0.7\,T_{\rm N}$ for $B=0$ ({\bf a}--{\bf c}) and $B=10$~Tesla ({\bf d}--{\bf f}). 
{\bf a} and {\bf d} show band dispersions with energy measured from the nominal Fermi level. 
The AHC $\sigma_{xy}$ is plotted as a function of the Fermi energy shift from the nominal value. 
{\bf b} and {\bf e} ({\bf c} and {\bf f}) show the Fermi surface on the $k_z = 0$ ($\pi/c$) plane. 
}
\label{fig:TN07}
\end{center}
\end{figure*}

While negative OHC below $T_{\rm N}$ agrees with our theoretical results {
(our low-temperature OHC $R_{\rm H} \sim -1.3 \times 10^{-9}$~m$^3$/C differs from the experimental value by a factor of $\sim 2$
without an adjustable parameter)}, 
the positive OHC above $T_{\rm N}$ does not--our theory with equal hole and electron mobilities predicts a near-zero value.  
This discrepancy may be resolved by incorporating a temperature-dependent carrier mobility into our analysis. 
However, our field-dependent AHC results offer an alternative interpretation of the observed OHC. 
As shown in Fig.~\ref{fig:AHC_OHC}~{\bf d} and Extended data Fig.~\ref{fig:AHC_vs_B}~{\bf a}, 
the AHC increases nearly linearly with magnetic field $B$ at low $B$, 
suggesting that a strongly positive $\sigma_{xy}/B$ at $T>0.85\,T_{\rm N}$ could enhance the low-field OHC $\rho_{yx}/B$. 
If hole and electron mobilities can be individually evaluated, 
the intrinsic Berry curvature contribution to the OHC may be isolated. 

Recent experiments by Wang et al. \cite{Wang2025} further support our findings. 
Their measurements of field- and temperature dependent Hall conductivity in FeTe reveal two components: 
one linear in magnetic field and another that is nonlinear.
The nonlinear component, extracted at zero field, is significantly enhanced below $T_{\rm N}$ 
and mirrors the temperature dependence predicted by our theory for $\sigma_{xy}$. 

{
Lastly, we comment on the orbital effect, that has been overlooked in our analysis. 
Although direct measurements of cyclotron mass in FeTe are lacking, 
it can be estimated via the Sommerfeld coefficient. 
Ma et al. reported DFT band masses of $4.665$~mJ/(mol\,K$^2$) (bicollinear antiferromagnetic state) and $4.783$~mJ/(mol\,K$^2$) (nonmagnetic state), both several times greater than the free electron mass \cite{Ma2009}. 
Assuming a cyclotron mass five times the free electron mass, the cyclotron frequency at $10$~Tesla is approximately 56 GHz (~0.2 meV or 2 K). Experimentally, Chen et al. found a much larger Sommerfeld coefficient of $34$~mJ/(mol\, K$^2$) \cite{Chen2009}. 
Because the lattice breaks Galilean invariance, correlation effects can enhance the cyclotron mass, 
like the Sommerfeld coefficient or quasiparticle effective mass. 
Thus, detecting orbital effects or Landau quantization requires temperatures well below $\sim 2$~K and high carrier mobility. 
}

Overall, our theoretical framework based on the spin-fermion model provides a microscopic explanation for the anomalous Hall response in FeTe, 
which cannot be be fully accounted for by the OHE alone. 
The transverse voltage in FeTe arises not only from the Lorentz force, but also from intrinsic deflections due to the Berry curvature, 
governed by the topological features of the band structure. 
This underscores the pivotal role of band topology in shaping the transport phenomena in FeTe. 

{
Field-induced AHE has been seen in antiferromagnetic metals like Mn$_3$Sn \cite{Li2023} 
as well as in non-magnetic Dirac semimetals, 
such as Cd$_3$As$_2$ \cite{Liang2017,Nishihaya2025} and ZrTe$_5$ \cite{Liang2018,Liu2021}. 
Our field-induced AHC reaches several hundred S/cm, which is 
comparable to or exceeding experimentally reported values for Dirac semimetals \cite{Nishihaya2025,Liu2021}. 
}
Our findings {
thus} establish FeTe as a prototypical platform for investigating intrinsic topological effects in collinear AFM systems. 
By elucidating the interplay between magnetism, topology, and electronic transport, this work deepens our understanding of the Hall effect  
and opens new avenues for exploring Berry curvature-driven effects in low-dimensional materials, 
with promising implications for spintronic and quantum technologies.

\section*{M\lowercase{ethods}}

\subsection*{DFT calculations and spin-fermion model}

Our theoretical analysis is based on a realistic spin-fermion model \cite{Liang2012}. 
As an advantage of this approach, one can investigate the electronic properties of the target system at finite temperature with an applied magnetic field. 
The band structure of itinerant electrons is derived via density functional theory calculations using Wannier90. 
For this purpose, we use the Vienna {\it ab} initio simulation package (VASP) \cite{Kresse1996a,Kresse1996b}, 
with the projector augmented wave method \cite{Blochl1994,Kresse1999} and the generalized gradient approximation in the parametrization of 
Perdew, Burke, and Enzerhof \cite{Perdew1996} for exchange-correlation. 
For both Fe and Te, standard potentials are used in the VASP distribution. 
We used an experimental structure at 80 K reported in Ref.~\cite{Li2009}. 
For an electronic self-consistent calculation in the paramagnetic phase, we use a $12\times12\times6$ $\mathbf k$-point grid and an energy cutoff of 500~eV. 
The spin-orbit coupling (SOC) is included, but the $+U$ correction is not included. 
Subsequently, we use the Wannier90 package \cite{Pizzi2020} to derive an effective tight-binding model projected onto the Fe $d$ states. 
The resulting tight-binding Hamiltonian is expressed as 
\begin{eqnarray}
H_d=\sum_{i,j} \sum_{a,b}\sum_{\sigma,\sigma'}{t_{ij}^{a\sigma,b\sigma'}d_{ia\sigma}^\dag}d_{jb\sigma'},
\end{eqnarray}
where $t_{ij}^{a\sigma,b\sigma'}$ is the hopping intensity between orbital $a$ and spin $\sigma$ at site $i$ and orbital $b$ and spin $\sigma'$ at site $j$, 
and $d_{i\alpha\sigma}^{\left(\dag\right)}$ is the annihilation (creation) operator of an Fe $d$ electron at site $i$, orbital $a$ and spin $\sigma$. 

Next, we carried out self-consistent calculations considering various magnetic configurations, including the bicollinear antiferromagnetic state 
(for details, see Supplementary Information, Note 2 and Fig. S6). 
Some of the antiferromagnetic orderings have a larger magnetic unit cell. 
Thus, we increase the size of the unit cell within the $xy$ plane to contain sixteen Fe sites and use a $6\times6\times6$ $\mathbf k$-point grid. 
To avoid possible spin canting, we turn off the SOC. 
As reported previously, the bicollinear antiferromagnetic state has the lowest energy. 
By mapping the energy landscape to that of a Heisenberg model $H_S=\sum_{l{\rm th\,neighbor}} \sum_{\langle ij \rangle_l} J_l \mathbf{S}_i \cdot \mathbf{S}_j$, 
we obtained exchange integrals up to the fourth neighbor sites as $J_1=5$, $J_2=12$, $J_3=8$, and $J_4=-7$~(meV), with $S=1$. 
Reflecting the edge-shared nature of FeTe$_4$ tetrahedra between the nearest-neighbor Fe sites \cite{Yin2011}, $J_1<J_2$. 
Within a mean-field approximation, this parameter set gives the N\'eel temperature of the bicollinear antiferromagnetic state as $T_N=\frac{8}{3}J_3 = 246$~K. 
We therefore renormalize $J_{1-4}$ so that $T_N=60$~K. 

Finally, a spin-fermion model is constructed by adding the coupling between localized spin $\mathbf{S}_i$ and itinerant Fe $d$ electrons to the Wannier tight-binding model such that 
its dispersion relation reproduces the DFT band structure in the bicollinear antiferromagnetic phase. 
Considering the tetragonal symmetry of FeTe, we introduced four exchange couplings as 
$H_{d-S}=\sum_{i,a,\sigma,\sigma'}{J_a \mathbf{S}_i \cdot \mathbf{\sigma}_{\sigma \sigma'}d_{ia\sigma}^\dag}d_{ia\sigma'}$. 
Here, $\mathbf{\sigma}$ is the Pauli matrix. 
We find that the DFT band structure near the Fermi level is reproduced using $J_{3z^2-r^2}=0.5$, $J_{x^2-y^2}=1.0$, $J_{xz}=J_{yz}=0.8$, $J_{xy}=0.7$~(eV) 
(see Supplementary Note 1. Fig.~S1~{\bf b}).

Using the spin-fermion model $H_d+H_S+H_{d-S}$, we analyze temperature and magnetic-field dependent magnetic structure, band structure, and anomalous Hall conductivity. 
The magnetic state at finite temperature and finite magnetic field is determined by $H_S-\sum_{i}{g\mu_B} \hat S_{iz}B$, where $g=2$ is the $g$ factor, $\mu_B$ is the Bohr magneton, 
and $B$ is a magnetic field applied perpendicular to the plane. 
We use a static mean-field approximation, i.e., 
$\mathbf{S}_i \cdot \mathbf{S}_j=\mathbf{S}_i \cdot \langle \mathbf{S}_j \rangle+\langle \mathbf{S}_i\rangle \cdot \mathbf{S}_j - \langle \mathbf{S}_i \rangle \cdot \langle\mathbf{S}_j\rangle$, 
and compute $\langle \mathbf{S}_i\rangle$ as $\langle S_{i \eta} \rangle ={\rm Tr} \hat S_{i \eta} \exp\bigl(-\hat{H}_i /T\bigr)/{\rm Tr} \exp\bigl( -\hat{H}_i/T\bigr)$, where $\hat{H}_i=\sum_{j \in l{\rm th\,neighbor}} J_l \mathbf{S}_i \cdot \langle \mathbf{S}_j\rangle-g \mu_B S_{iz}B$ and $\hat S_{i\eta}$is a $3\times3$ matrix for the $\eta$ component of spin $S=1$. 
Here, we assume nonzero spin anisotropy so that $\langle \mathbf{S}_j\rangle$ is in the $yz$ plane. 
The magnetic state then influences the electronic band structure via $H_{d-S}$. 

\subsection*{Berry curvature and anomalous Hall conductivity}
After solving the fermionic part of the Hamiltonian, the band-dependent Berry curvature is computed using~\cite{Wang2006}
\begin{eqnarray}
\Omega_{xy}^\alpha\left(\mathbf{k}\right) 
= -2 \, {\rm Im} \! \sum_{\beta (\neq \alpha)} \! \frac{\left\langle\alpha\mathbf{k}|{\hat{v}}_x\left(\mathbf{k}\right)|\beta\mathbf{k}\right\rangle
\left\langle\beta\mathbf{k}|{\hat{v}}_y\left(\mathbf{k}\right)|\alpha\mathbf{k}\right\rangle}
{\left(\varepsilon_{\alpha\mathbf{k}}-\varepsilon_{\beta\mathbf{k}}\right)^2}.
\end{eqnarray}
{
Here, $|\left.\alpha\mathbf{k}\right\rangle$ is the Bloch eigenstate associated with band $\alpha$ at momentum $\bf k$. }
${\hat{v}}_\xi\left(\mathbf{k}\right)$ is the velocity operator matrix along the $\xi \left(=x,y\right)$ direction given by 
${\hat{v}}_\xi\left(\mathbf{k}\right)=\frac{\partial\hat{H}\left(\mathbf{k}\right)}{\hbar\partial k_\xi}$, where 
$\hat{H}\left(\mathbf{k}\right)$ is the momentum Fourier transform of the fermionic Hamiltonian $H_d$. 
The anomalous Hall conductivity $\sigma_{xy}^A$ is then computed via
\begin{eqnarray}
\sigma_{xy}^A=-\frac{q_e^2}{\hbar N V_c}\sum_{\alpha,\mathbf{k}}{\Omega_{xy}^\alpha\left(\mathbf{k}\right)}f\left(\varepsilon_{\alpha \mathbf{k}}-E_{\rm F}\right),
\end{eqnarray}
where 
$V_c$ is the volume of the magnetic unit cell, and $q_e$ is the electron charge.
For the momentum integration, we used a $100\times200\times200$ $\mathbf k$-point grid.

\section*{D\lowercase{ata availability}}
All data that support the findings of this study are available at \url{https://doi.org/10.5281/zenodo.19665117}. 

\section*{C\lowercase{ode availability}}
The Ab initio calculations are carried out using the code VASP, and a realistic tight-binding model is constructed using Wannier90. 
Other calculations using a spin-fermion model are done with in-house codes that are available from the corresponding author upon reasonable request.

\section*{A\lowercase{cknowledgments}}
The research by S.O. and A.M. was supported by the U.S. Department of Energy, Office of Science, Basic Energy Sciences, Materials Sciences and Engineering Division. 
N.N. was supported by JSPS KAKENHI Grant Numbers 24H00197, 24H02231 and 24K00583, and by the RIKEN TRIP initiative. 
S.S.P.P. was funded by the European Union (FUNLAYERS, project number 101079184).
S.O. and A.M. thank the authors of Ref.~\cite{Wang2025} for sharing their experimental data prior to publication 
and their fruitful discussions.

\section*{A\lowercase{uthor contributions}}
S.O. and S.S.P.P. conceived and designed this project. 
S.O. developed a theoretical model and carried out numerical calculations. 
A. M., N. N., and S.S.P.P contributed the theoretical analysis. 
All authors discussed the results and prepared the manuscript.

\section*{C\lowercase{ompeting interests}}
The authors declare no competing interests.


\section*{E\lowercase{xtended data}}

%
%
%

\clearpage

\def\bibsection{\section*{S\lowercase{upplementary} R\lowercase{eferences}}} 

\setcounter{MaxMatrixCols}{10}




\onecolumngrid

\newpage
\begin{center}
{\large \bf Supplementary Information: Berry curvature and field-induced intrinsic anomalous Hall effect in an antiferromagnet FeTe}\\
\vspace{1em}
Satoshi Okamoto,$^1$ Adriana Moreo,$^{1,2}$ Naoto Nagaosa,$^{3,4}$ and Stuart S. P. Parkin$^5$ \\
\vspace{0.5em}
{\small \it $^1$Materials Science and Technology Division, Oak Ridge National Laboratory, Oak Ridge, Tennessee 37831, USA}\\
{\small \it $^2$Department of Physics and Astronomy, The University of Tennessee, Knoxville, Tennessee 37996, USA}\\
{\small \it $^3$RIKEN Center for Emergent Matter Science (CEMS), Wako, Saitama 351-0198, Japan}\\
{\small \it $^4$Fundamental Quantum Science Program (FQSP), TRIP Headquarters, RIKEN, Wako, Saitama 351-0198, Japan}\\
{\small \it $^5$Max Planck Institute of Microstructure Physics, Halle(Saale) 06120, Germany}
\end{center}

\renewcommand{\thetable}{S\Roman{table}}
\renewcommand{\thefigure}{S\arabic{figure}}
\renewcommand{\thesection}{Supplementary Note \arabic{section}}
\renewcommand{\thesubsection}{\arabic{subsection}}
\renewcommand{\thesubsubsection}{\arabic{subsection}.\arabic{subsubsection}}
\renewcommand{\theequation}{S\arabic{equation}}

\setcounter{secnumdepth}{3}

\setcounter{equation}{0}
\setcounter{figure}{0}

\section{Theoretical consideration of bicollinear order and band structure}
\begin{figure}[h]
\begin{center}
\includegraphics[width=0.7\columnwidth, clip]{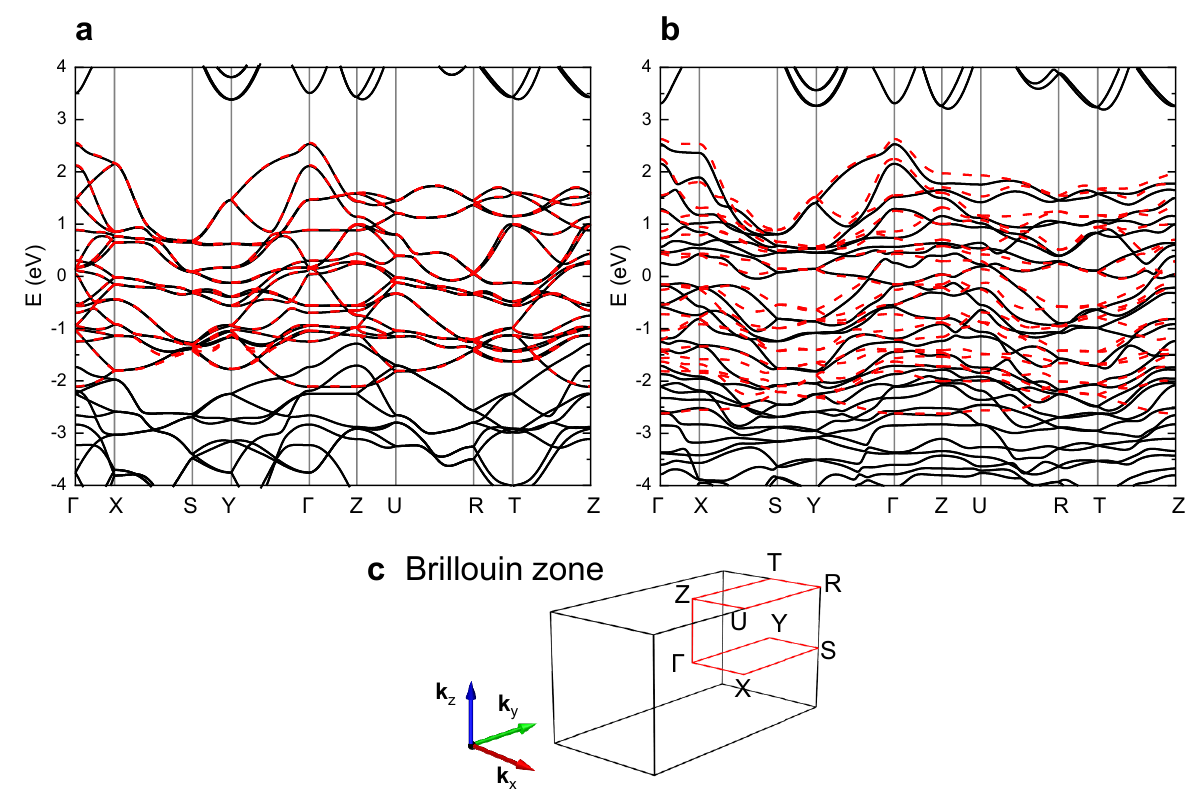}
\caption{Band structure of FeTe. {\bf a}. DFT band structure (black lines) and Wannier interpolations (red dashed lines) for a nonmagnetic state. 
{\bf b}. DFT band structure (black lines) and spin-fermion model (red dashed lines) for a bicollinear antiferromagneic state. 
{\bf c}. Magnetic Brillouin zone used for a and b. Note that the reduced zone is also used for the nonmagnetic state.}
\label{fig:DFTband}
\end{center}
\end{figure}

Here, we provide the details of theoretical calculations. 
Fig.~\ref{fig:DFTband}~{\bf a} shows the DFT band structure (black solid lines) of FeTe without magnetic ordering. 
Wannier interpolation projected onto Fe $d$ states is also plotted as red dashed lines, showing an excellent agreement. 
Fig.~\ref{fig:DFTband}~{\bf b} compares the DFT band structure (black solid lines) and dispersion relation of the spin-Fermion model in the bicollinear AFM state. 
Since the spin-fermion model is constructed using the Wannier tight-binding model obtained in the paramagnetic state, the agreement between the two results is not perfect. 
Nevertheless, the two sets of curves show reasonable agreement near the Fermi level $E=0$. 
Away from the Fermi level, especially below $E \lesssim -2$~(eV), the spin-fermion model does not reproduce the DFT band structure, 
because it does not include Te $p$ states.  

\begin{figure}[h]
\begin{center}
\includegraphics[width=0.7\columnwidth, clip]{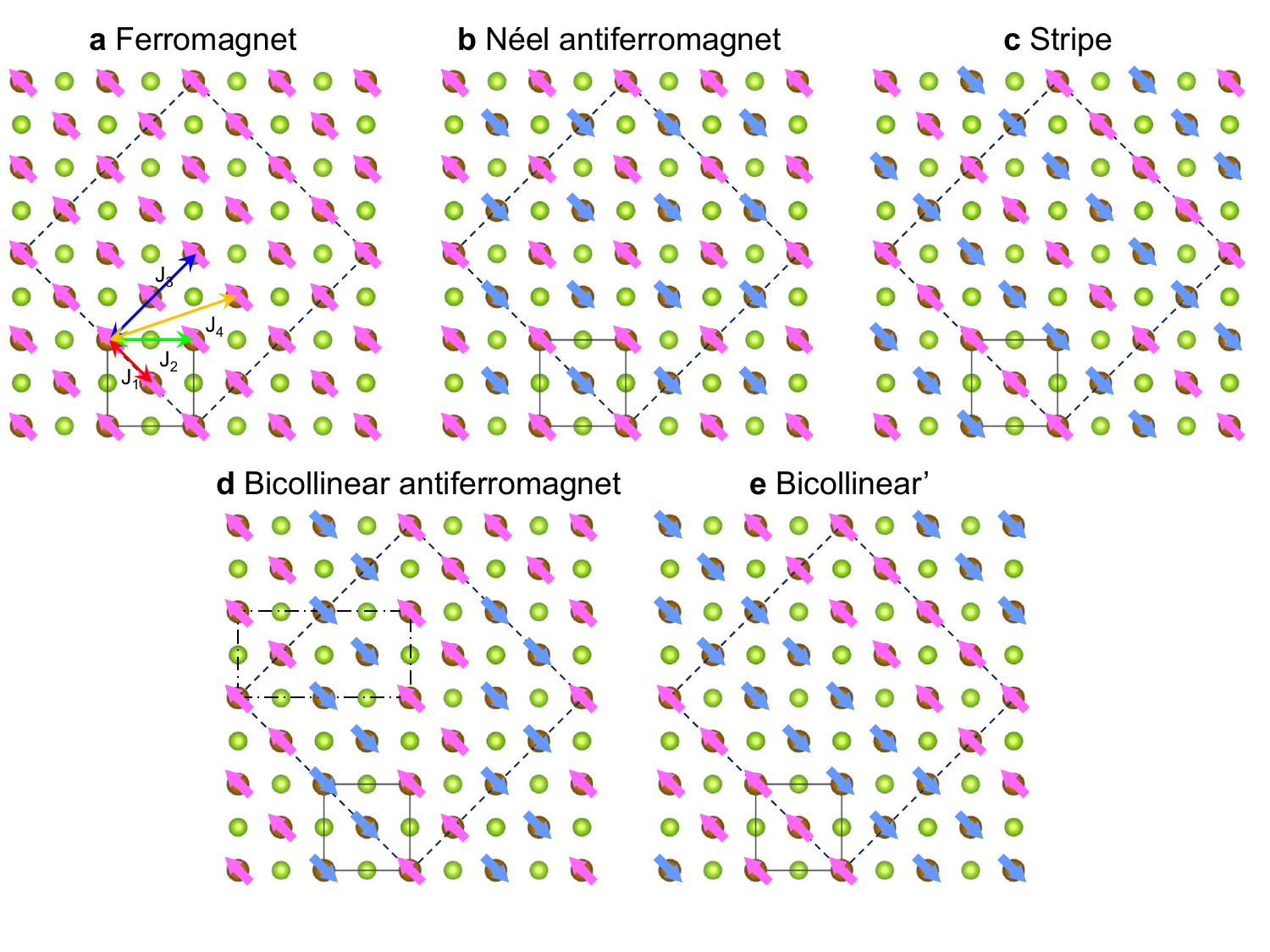}
\caption{Magnetic ordering considered to construct a localized spin model. 
Small squares shown with thin solid lines are the original unit cell. Large squares with dashed lines are the unit cells considered in DFT calculations. 
Rectangular with dash-dotted lines in d is the magnetic unit cell for the bicollinear antiferromagnet. }
\label{fig:magneticorder}
\end{center}
\end{figure}

To extract exchange interactions between localized spins, we carry out DFT calculations considering five magnetic ordering as shown in Fig.~\ref{fig:magneticorder}. 
The resulting total energy is summarized in Table~\ref{table:energy}, confirming that the bicollinear antiferromagnetic state is the most stable. 
The ordered Fe moment is found to be about $2\mu_B$ for all the cases. 
Thus, assuming that the Fe moment is constant $S=1$, we map the energy landscape to that of the Heisenberg model 
$H_S=\sum_{l{\rm th\,neighbor}} \sum_{\langle ij \rangle_l} J_l \mathbf{S}_i \cdot \mathbf{S}_j$. 
For magnetic ordering as shown in Fig.~\ref{fig:magneticorder}, we find 
\begin{eqnarray}
E_{\mathrm{FM}} \!\!&=&\!\! E_0+32J_1S^2+32J_2S^2+32J_3S^2+64J_4S^2, \\
E_{\mathrm{N\'eel}}\!\!&=&\!\!E_0-32J_1S^2+32J_2S^2+32J_3S^2-64J_4S^2, \\
E_{\rm Stripe}\!\!&=&\!\!E_0-32J_2S^2+32J_3S^2, \\
E_{\rm Bicollinear}\!\!&=&\!\!E_0-32J_3S^2, \\
E_{\rm Bicollinear'}\!\!&=&\!\!E_0+16J_1S^2-32J_4S^2, 
\end{eqnarray}
where, $E_0$ is a constant. 
From this set of total energy, we find $J_1=5$, $J_2=12$, $J_3=8$, and $J_4=-7$~(meV).

\begin{table}
\begin{center}
\caption{DFT total energy in the unit of eV of five magnetic states considered. 
The structural cell contains sixteen Fe sites as shown in Fig.~\ref{fig:magneticorder}.}
\begin{tabular}{ c | c |  c | c | c}
\hline
 Ferromagnet & N\'eel AFM & Stripe & Bicollinear & Bicollinear' \\ 
\hline \hline
 $-184.184$ & $-183.590$ & $-184.676$ & $-184.815$ & $-184.238$\\
\hline
\end{tabular}
\label{table:energy}
\end{center}
\end{table}

\section{Temperature and field dependent magnetization}

\begin{figure}[h]
\begin{center}
\includegraphics[width=1.0\columnwidth, clip]{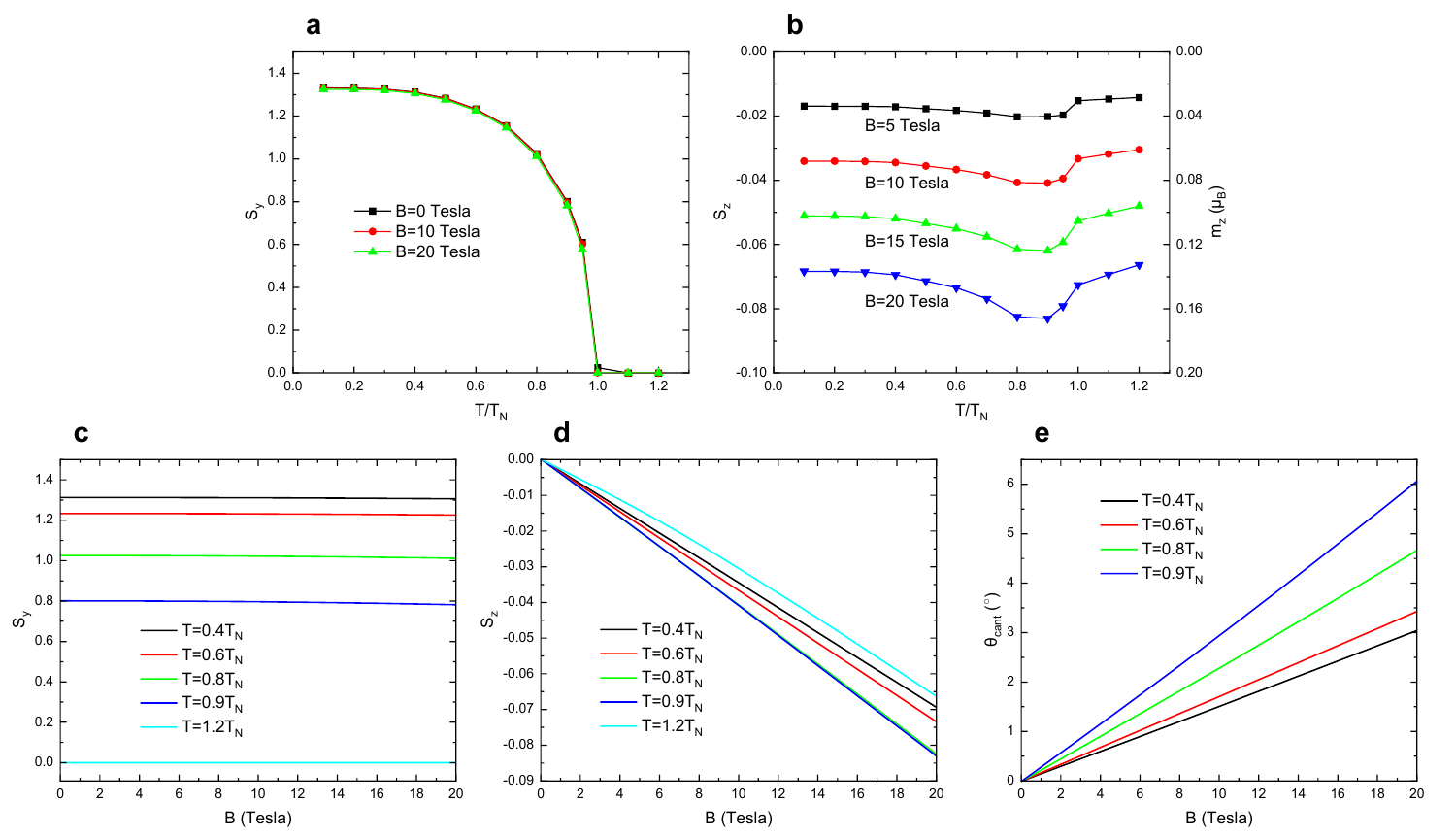}
\caption{Temperature and field dependent magnetization of Fe $d$ electrons. 
{\bf a} and {\bf b} show the temperature dependent staggered magnetization $S_y$ and the longitudinal magnetization $S_z$, respectively, under various magnetic field strengths.
{\bf c} and {\bf d} show the field dependent $S_y$ and $S_z$, respectively, at various temperatures. 
Resulting canting angle $\theta_{\rm cant} = \tan^{-1} (-S_z/S_y)$ is shown in {\bf e}.} 
\label{fig:bandmagne}
\end{center}
\end{figure}

The magnetization of Fe $d$ electrons, as influenced by temperature and magnetic field, is analyzed using the spin-fermion model. 
Utilizing Bloch eigenstates, the staggered magnetization component ($S_y$) and the longitudinal magnetization along the field direction ($S_z$) are computed by 
\begin{eqnarray}
S_y = \frac{1}{4N} \frac{1}{2}
\sum_{i, a, \sigma} \sum_{\alpha, \mathbf{k}} \langle \alpha \mathbf{k} |
d^\dag_{i a \sigma} \eta_i \sigma^y_{\sigma \bar \sigma} d_{i a \bar \sigma} 
|\alpha \mathbf{k}\rangle, \\
S_z = \frac{1}{4N} \frac{1}{2}
\sum_{i, a, \sigma} \sum_{\alpha, \mathbf{k}} \langle \alpha \mathbf{k} |
d^\dag_{i a \sigma} \sigma^z_{\sigma \sigma} d_{i a \sigma}
|\alpha \mathbf{k}\rangle. 
\end{eqnarray}
Here, the index $i$ ranges from 1 to 4 within the magnetic unit cell, $\eta_i$ assigned as $+1$ or $-1$ depending on the direction of the staggered magnetization, and 
$\mathbf{\sigma}^{\nu}$ denotes the $\nu$ component of the Pauli matrices. 
Figure~\ref{fig:bandmagne} summarizes the magnetization results, where panels {\bf a} and {\bf b} display $S_y$ and $S_z$ as functions of temperature 
under various magnetic field strengths. 
$S_y$ remains unaffected by the magnetic field $B$, whereas $S_z$ exhibits a strong dependence on it. 
However, the induced $S_z$ is relatively small, resulting in a maximum induced magnetic moment along the field ($m_z = - g \mu_B S_z$) of approximately $0.16 \mu_B$ for $B$ up to 20~Tesla. 
The weak field sensitivity of $S_y$ and pronounced field sensitivity of $S_z$ are further illustrated in panels {\bf c} and {\bf d} of Fig.~\ref{fig:bandmagne}.
The canting angle $\theta_{\rm cant}$ of magnetic moment varies almost linearly with the applied field, 
mirroring the field dependence of $S_Z$, as shown in Fig.~\ref{fig:bandmagne}~{\bf e}.

\section{Ordinary Hall effect}
Similar to AHC, ordinary Hall coefficient (OHC) is computed using the spin-Fermion model. 
We consider the classical low-field Hall coefficient and follow the procedure outlined in Ref.~\cite{SYates2007}. 
Here, the OHC is given by 
\begin{equation}
R_H=\frac{\rho_{xy}}{B}=\frac{\sigma_{xy}^O}{\sigma_{xx} \sigma_{yy}}, 
\end{equation}
where $\sigma_{\xi\xi}$ is the longitudinal conductivity given by
\begin{eqnarray}
\sigma_{\xi\xi}=q_e^2\frac{1}{N}\sum_{\alpha,\mathbf{k}}{\tau_\alpha^2\left(\mathbf{k}\right)v_{\alpha,\xi}^2\left(\mathbf{k}\right)}\left\{-\frac{\partial f\left(\varepsilon_{\alpha\mathbf{k}}\right)}{\partial\varepsilon_{\alpha\mathbf{k}}}\right\},
\end{eqnarray}
and $\sigma_{xy}^O$ is the $B$-linear coefficient of low-field Hall conductivity given by
\begin{eqnarray}
\sigma_{xy}^O=q_e^3\frac{1}{N}\sum_{\alpha,\mathbf{k}}{\tau_\alpha^3\left(\mathbf{k}\right)}\left\{v_{\alpha,x}^2\left(\mathbf{k}\right)\ \eta_{\alpha,yy}\left(\mathbf{k}\right)-v_{\alpha,x}\left(\mathbf{k}\right)\ v_{\alpha,y}\left(\mathbf{k}\right)\ \eta_{\alpha,xy}\left(\mathbf{k}\right)\right\}\left\{-\frac{\partial f\left(\varepsilon_{\alpha\mathbf{k}}\right)}{\partial\varepsilon_{\alpha\mathbf{k}}}\right\}.
\end{eqnarray}
Here, $v_{\alpha,\xi}\left(\mathbf{k}\right)$ and $\eta_{\alpha,\xi\zeta}\left(\mathbf{k}\right)$ are given by 
\begin{eqnarray}
v_{\alpha,\xi}\left(\mathbf{k}\right)=\langle\alpha\mathbf{k} |\partial_\xi\hat{H}\left(\mathbf{k}\right) |\alpha\mathbf{k} \rangle
\end{eqnarray} 
and 
\begin{eqnarray}
\eta_{\alpha,\xi\zeta}\left(\mathbf{k}\right)
=\langle\alpha\mathbf{k} |\partial_\xi\partial_\zeta\hat{H}\left(\mathbf{k}\right) |\alpha\mathbf{k} \rangle 
+\frac{\langle\alpha\mathbf{k}|\partial_\zeta\hat{H}\left(\mathbf{k}\right)|\beta\mathbf{k}\rangle}{\varepsilon_\alpha\left(\mathbf{k}\right)-\varepsilon_\beta\left(\mathbf{k}\right)} \langle\beta\mathbf{k}|\partial_\xi\hat{H}\left(\mathbf{k}\right)|\alpha\mathbf{k} \rangle+\langle\alpha\mathbf{k}|\partial_\xi\hat{H}\left(\mathbf{k}\right)|\beta\mathbf{k} \rangle 
\frac{\langle\beta\mathbf{k}|\partial_\zeta\hat{H}\left(\mathbf{k}\right)|\alpha\mathbf{k}\rangle}{\varepsilon_\alpha\left(\mathbf{k}\right)-\varepsilon_\beta\left(\mathbf{k}\right)}, 
\end{eqnarray}
respectively, with $\partial_\xi=\frac{\partial}{\hbar\partial k_\xi}$, and $\tau_\alpha\left(\mathbf{k}\right)$ is the carrier lifetime. 
{
For the numerical result shown in Fig. 2 {\bf c}}, we carry out the momentum integration using 
a $100\times200\times200$ $\mathbf k$-point grid, assuming that $\tau_\alpha\left(\mathbf{k}\right)$ is independent of the band index $\alpha$ and momentum $\mathbf k$.

\end{document}